\documentclass{article}

\usepackage{arxiv}

\usepackage[utf8]{inputenc} 
\usepackage[T1]{fontenc}    
\usepackage{hyperref}       
\usepackage{url}            
\usepackage{booktabs}       
\usepackage{amsfonts}       
\usepackage{nicefrac}       
\usepackage{microtype} 
\usepackage{longtable}
\usepackage{graphicx}
\usepackage{doi}
\usepackage{subfigure} 
\usepackage{ragged2e}
\usepackage{pdflscape}
\usepackage{graphicx}
\usepackage{caption}
\usepackage{amsmath}
\usepackage{multirow}
\usepackage{tabularx}
\usepackage{amsfonts}
\usepackage{amssymb}
\usepackage{graphicx}
\usepackage{color}

\usepackage{amsmath}
\usepackage{cite}
\usepackage{amsfonts,amsmath}
\usepackage{amssymb, nccmath}
\usepackage{algorithm}
\usepackage{algorithmic}
\usepackage{amssymb, nccmath}
\usepackage{algorithm}
\usepackage{algorithmic}
\usepackage{dsfont}
\usepackage{booktabs}
\usepackage[utf8]{inputenc}
\usepackage{dsfont}
\usepackage{colortbl, hhline}
\usepackage{booktabs}
\usepackage{graphicx}
\usepackage{grffile}
\usepackage{grffile}
\usepackage{array} 
\usepackage{mathtools}
\usepackage{fontenc}
\usepackage{tikz}

\usepackage{afterpage}
\definecolor{yellow}{HTML}{E0B405}
\definecolor{timeLineProgressColor}{HTML}{E0B405}
\definecolor{currentPosition}{HTML}{A0C4A5}
\definecolor{gold}{HTML}{CA4F04}
\definecolor{beige}{rgb}{0.96, 0.96, 0.86}
\definecolor{frameBG}{HTML}{F5F5DC}
\definecolor{champagne}{rgb}{0.97, 0.91, 0.81}
\definecolor{apricot}{rgb}{0.98, 0.81, 0.69}
\definecolor{bronze}{rgb}{0.8, 0.5, 0.2}

\title{An Intelligent Security centered Resource-Efficient Resource Management Model  for Cloud Computing Environments}

\author{ \href{https://orcid.org/0000-0002-9689-6387}{\includegraphics[scale=0.06]{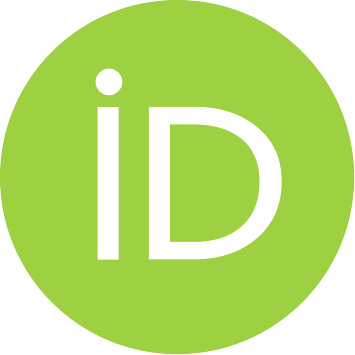}\hspace{1mm}Deepika Saxena}\thanks{The authors would like to thank National Institute of Technology Kurukshetra, India for financially supporting this research work.} \\
	Department of Computer Applications\\
    National Institute of Technology\\
	Kurukshetra, India \\
	\texttt{deepika\_6180096@nitkkr.ac.in} \\
	\And
	\href{https://orcid.org/0000-0002-8053-5050}{\includegraphics[scale=0.06]{orcid.pdf}\hspace{1mm}Ashutosh Kumar Singh} \\
		Department of Computer Applications\\
      National Institute of Technology\\
	Kurukshetra, India \\
	\texttt{ashutosh@nitkkr.ac.in} }

\renewcommand{\shorttitle}{\textit{Predictive Resource Management for Cloud Environments}}

\hypersetup{
pdfsubject={q-bio.NC, q-bio.QM},
pdfauthor={David S.~Hippocampus, Elias D.~Striatum},
pdfkeywords={First keyword, Second keyword, More},
}

\begin{document}
\maketitle

\begin{abstract}
This paper proposes a conceptual model for a secure and performance efficient workload management model in cloud environments. In this model, a resource management unit is employed for energy and performance proficient allocation of virtual machines while ensuring the secure processing of users’ applications by defending against the data breaches due to unauthorized access of virtual machines in real-time. The resource management unit is guided by a secure virtual machine management unit which is designed to generate information regarding unauthorized access or inter-communication links among active virtual machines. Also, a workload analyser unit operates concurrently to estimate resource utilization information to assist the resource management unit in performance efficient allocation of virtual machines. Contrary to prior works which engages access control mechanisms, and encryption and decryption of data before transfer, and use of tunnelling for prevention of an unauthorised access to virtual machines which raises excess computational cost overhead, the proposed  model  operates diversely for efficiently serving the same purpose.

\end{abstract}


\section{Introduction}
Cloud Computing enabled with seamless pool of hardware and software resources with advanced computation and numerous benefits viz., service availability, reliability, elasticity, and maximum scalability at minimum upfront investment has driven business enterprise, scientific and research organisations towards the cloud datacentres \cite{singh2021quantum, saxena2014review, saxena2015ewsa, saxena2015highly, saxena2016dynamic, singh2019secure, singh2018data,  singh2021cryptography, singh2020online,kumar2009vmanage, singh2022privacy,  kumar2016dynamic}. The users requests or applications are executed on different virtual machines hosted on different severs within the datacentre \cite{kumar2017mppt, kumar2018long,gupta2022holistic,  kumar2018workload,    kumar2020biphase, chhabra2016dynamic,varshney2022plant,kumar2019cloud,chhabra2018oph,patel2021lstm, chhabra2018probabilistic, chhabra2019dynamic,kumar2020adaptive,gupta2019confidentiality, chhabra2019optimal, gupta2018probabilistic,sasamal2018efficient,  kumar2020cloud}. The cloud virtualization technology facilitates elasticity for satisfying dynamically varying resource demands of users by allocating and deallocating required number of virtual machines on the shared physical servers \cite{godha2021flooding,kumar2020decomposition, singh2019sql,kumar2020ensemble, kumar2021discussion, gupta2019dynamic, gupta2019layer, gupta2020framework, saxena2018abstract, saxena2020auto,singh2022privacysmart}. However, the sharing of physical resources and processing of confidential applications on the third-party cloud servers prompts a biggest concern of ‘security’ among all the cloud users \cite{ gupta2022tidf, saxena2020communication, gupta2022compendium, raj2016low, gupta2020seli,saxena2020security, gupta2022hisa, saxena2021energy, gupta2022pca, saxena2021op,gupta2022mlrm, devkota2018image, saxena2021osc, gupta2022differential, saxena2021proactive,gupta2022auxiliary,  saxena2021securevmp, gupta2020guim, makkar2022secureiiot, gupta2020integrated, gupta2020mlpam,  gupta2021data,   gupta2022privacy, singh2020online}. 

 The virtual machines dedicated for the processing of applications of a user needs to communicate for data exchange among them \cite{swain2022efficient, yu2016stochastic,  saxena2020communication, kumar2021resource,gaur2022efficient, kumar2021self, singh2022privacy1,nader2015designing, singh2019stock, kaur2017comparative, kumar2021performance,  gaur2021testable,saxena2021survey}. Most of the data hacks and security attacks occur by reason of data transfer leveraging the security loopholes or vulnerabilities of compute, storage, and network devices within datacentre \cite{choudhary2021review, makaju2018lung, patel2021review, saxena2022vm, gupta2022differential22, varshney2021machine,yang2018emotion,  saxena2022high}. Also, the co-location of virtual machines of multiple users on a common physical server often leads to malicious activities like, data hampering and leakage of cloud user’s sensitive information \cite{ge2016evaluating, chhabra2020security,goh2015comprehensive, chhabra2021dynamic, sharma2021lightweight,tripathi2020review, chauhan2020survey, sharma2019fast, yadav2019advancements,kaur2017data,singh2018data, gupta2022quantum}. A malicious user or attacker may initiate one or group of virtual machines and exploit multiple network routes to attain co-residency with the target virtual machine \cite{tagliacane2016network, kesharwani2021real,acharya2021host,pradhan2021comparative, tiwari2021credit, patel2021lstm, jalwa2021comprehensive,hura2020advances,deepika2020review, gupta2022differential55, gupta2022differential77}. The vulnerability and susceptibilities of hypervisor or virtual machine management layer allow malicious user to easily compromise all the virtual machines hosted on it. Therefore, the key challenge for cloud datacentre is securing user data distributed on the shared physical servers while maintaining scalability of resources along with high performance and cost optimization \cite{leng2012link,agarwal2019authenticating,godha2019architecture, chhabra2020secure, islam2012empirical, wang2013new, sharma2019failure, beloglazov2012optimal, tripathihedcm}. 
 
 The existing approaches furnished for securing real-time data communications among virtual machines and other different entities includes encryption and decryption of data before transfer, use of virtual private networks, tunnelling, and use of passwords, captchas for prevention of an unauthorised access to virtual machines of benign user etc. Also, minimization of the physical server sharing and periodic migration of virtual machines are some practices employed for reduction of security attacks and safeguarding online data execution and communication. Despite the adoption of aforementioned security upgrading measures, the misconfiguration and mismanagement at cloud platform is the topmost cause for leakage of terabytes of sensitive data of millions of cloud users across the world. Hence, a robust security mechanism during user data processing and communication at virtual machine placement and physical resource management level is needed to combat the aforementioned bottleneck problem of security threats in cloud environments. 
 \section{Proposed Model}
 
\begin{figure}[!htbp]
 	\centering
 	\includegraphics[width=1.0\linewidth]{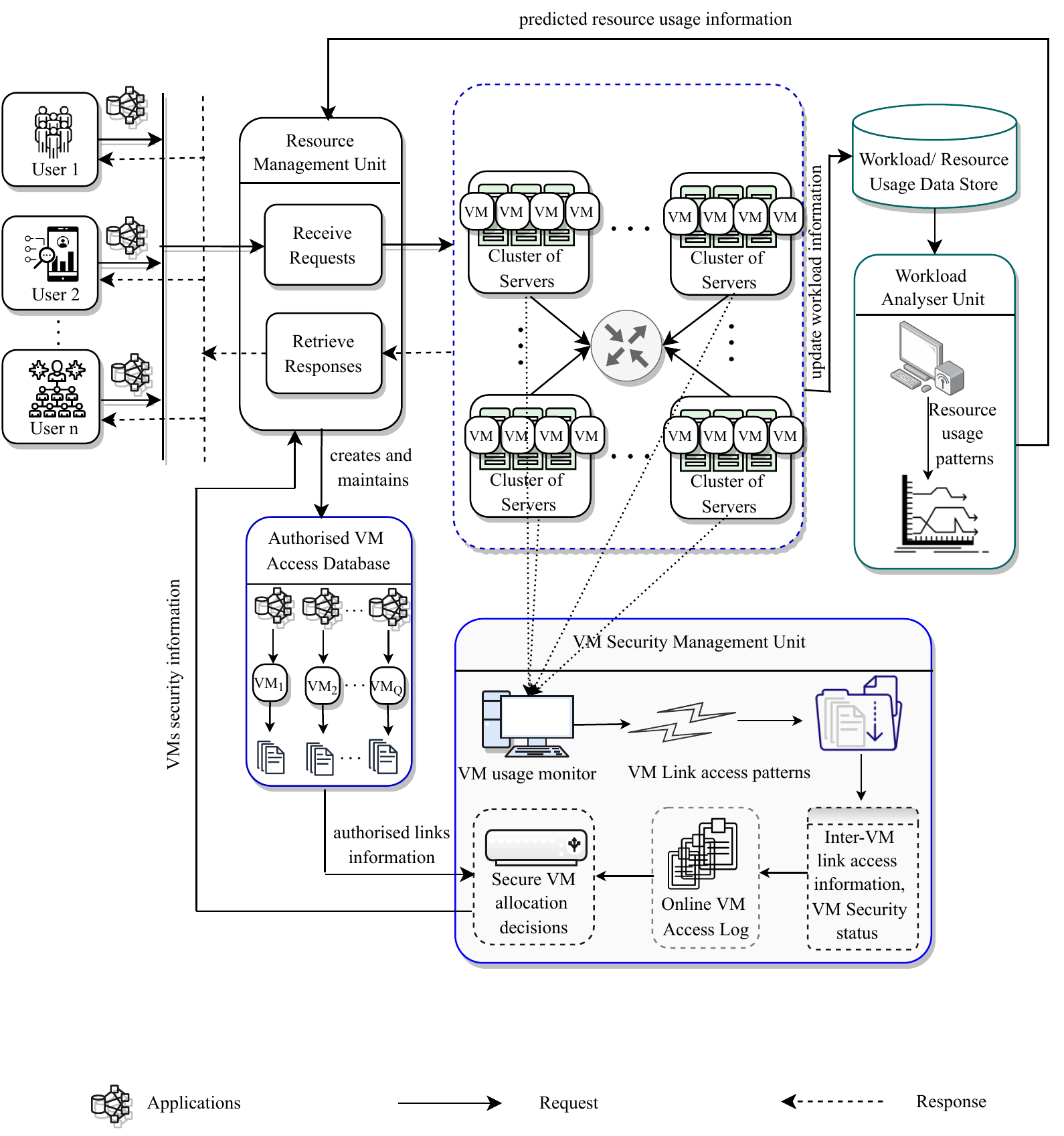}
 	\caption{Proposed model architectural overview }
 	\label{fig:fig1}
 \end{figure}

 This section discusses an overview of the resource distribution and allocation of virtual machines for securing data execution and unauthorized access within the cloud datacenter in a simplified format that are further explained in the detailed description of the paper. The proposed conceptual model addresses the challenges of security breaches, virtual machine hijack and data leakage, network security attacks, unauthorized data access along with performance degradation due to server under/over-utilization by proposing a novel virtual machine security ingrained cloud resource allocation model. The architectural view of proposed model is illustrated in Fig. \ref{fig:fig1}.
 
A resource management unit is employed to receive application execution requests from numerous users across the globe. These applications maybe received in any possible format such as high-performance computing applications, scientific workflows, military, medicinal or research-based applications, online business and marketing etc. The resource management unit divides these applications into one or more components or tasks to be executed concurrently on different virtual machines.  All the virtual machines engaged in processing of the component of a common application needs to communicate and exchange information with each other builds a virtual network of virtual machines with authorized access links. The resource management unit store and maintains this information of authorized virtual machine links in an Authorized Virtual machine Access Database (AVAD).  \par
 The resource management unit assigns different tasks to distinct virtual machines deployed on various servers within the datacenter. During execution of tasks by different virtual machine, the resources viz., CPU, memory, disk, bandwidth utilization are recorded and Workload/Resource Usage Data Store is created and updated periodically. The information saved in this data store is provided as input for training and re-training of a machine learning based workload analyser. This workload analyser is employed to estimate the future resource utilization on different virtual machines and servers. The resource management unit utilizes estimated resource usage information in making various proactive decisions confined to load management such as creation, termination, or migration of different virtual machines, handling of over/under-load conditions of servers etc., for an efficient management of resources.
 The proposed model employed a Virtual Machine (VM) Security Management Unit to ensure secure execution of user application on group of virtual machines hosted on same or different servers. This unit allows monitoring of inter-communication links among active virtual machines in real-time and enable detection of attacker virtual machines for mitigation of security breaches within the datacenter. During task execution, the resource usage is collected to update workload data store and inter-communication links are monitored and stored in Current VM Access Links (CVAL) log.  All the live inter-communication links among virtual machines are maintained in a log which helps to determine unauthorized access links and terminate attacker/malicious virtual machines before the occurrence of any security breaches. The proposed model is capable avoiding data security attacks because of new/unknown as well as known attacker virtual machine. 
  \par
 The present approach illustrates  a general cloud computing environment accompanying role of management of resources within cloud environment wherein, cloud infrastructure  comprises of clusters of servers  connected via router networking device for communicating with one another as per the requirement. Each cluster comprises of numerous physical server machines of same or varying resource capacity. The resources may include Central Processing Unit (CPU), RAM, Disk memory, and bandwidth etc. The virtual machines  which maybe in the form of computing, storage or networking instances are hosted on servers for processing of cloud user application requests. 
 The cloud users \{$User $ 1, $User$ 2, ..., $User $ n\} may send application requests to a Resource Management Unit for processing at the cloud infrastructure. The Resource Management Unit receives requests in the form of applications from various cloud users \{$User $ 1, $User $ 2, ..., $User $ n\}. Moreover, it facilitates the virtual machine management including virtual machine scheduling, placement, migration, and handling over/under-loading conditions by reason of inefficient resource distribution among virtual machines. Further, it serves tasks management by performing the function of user application division into sub-components or tasks, assignment and scheduling of these tasks’ execution on different virtual machines. The output processed tasks belonging to an application are integrated to generate the response for the user retrieved via Resource Management Unit. 
 
 \section{Secure Resource Management}
\begin{figure}[!htbp]
 	\centering
 	\includegraphics[width=1.0\linewidth]{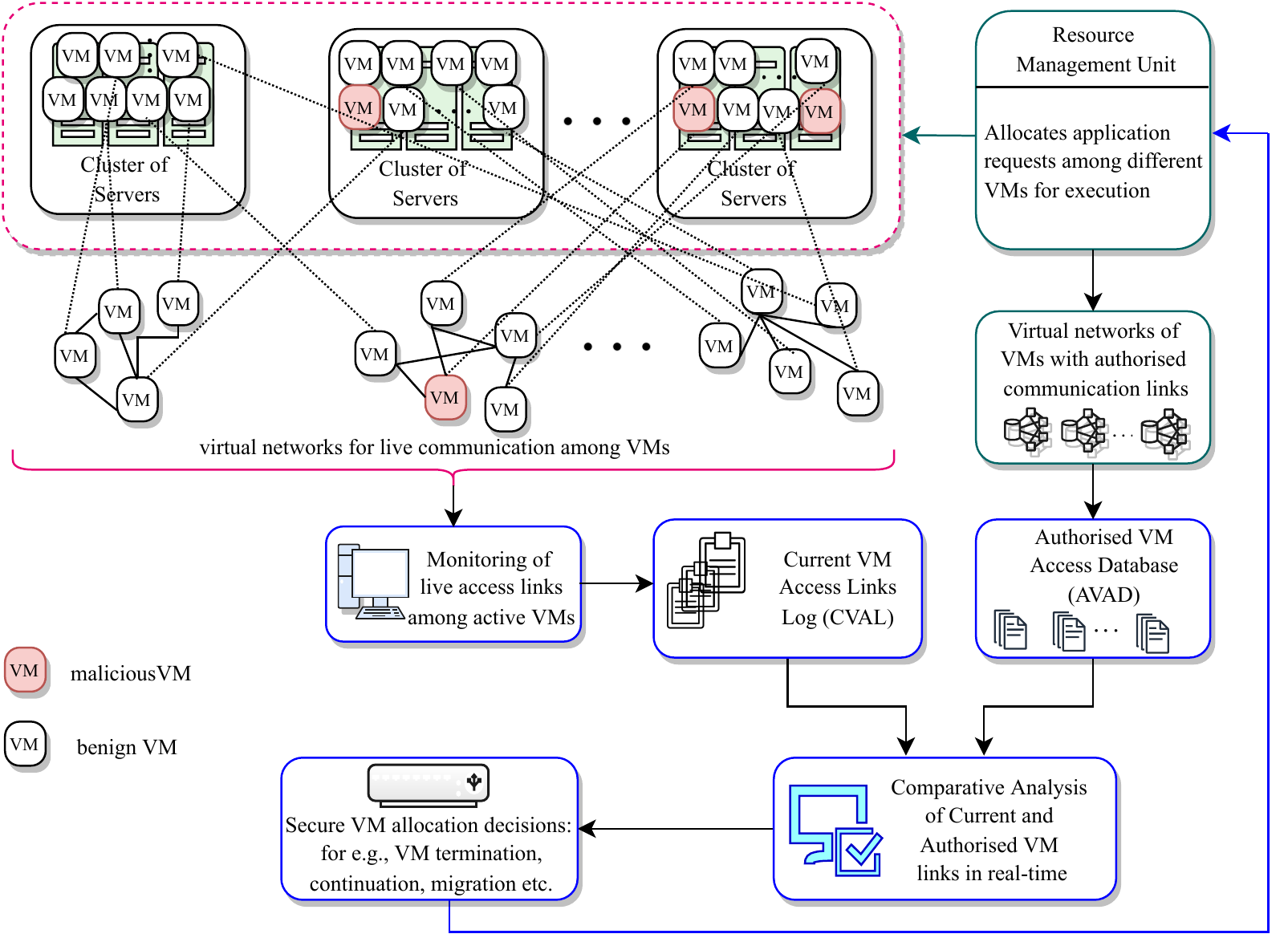}
 	\caption{Security focused resource management unit}
 	\label{fig:fig3}
 \end{figure}

 Fig. \ref{fig:fig3} portrays a complete view of a secure and performance efficient load management model in cloud environment in accordance with the embodiment of the proposed model. The cloud users leverages services of cloud infrastructure by sending requests in the form of different applications to be received by the resource management unit. These applications are divided into tasks in accordance with the tasks management, which are assigned for execution to different virtual machines subject to the resource requirement of a task is lesser than or equals to the resource capacity of virtual machine in accordance with the virtual machine management. The virtual machines utilize physical resource capacity of the servers upon which they are hosted. 
 The resource utilization or received workload information by different virtual machines on different servers are recorded in a Workload/Resource Usage Data Store. This information is used for periodic training and re-training of a machine learning algorithm based Workload Analyser Unit which is further described in detail in accordance with an embodiment of the proposed model as depicted in Fig. \ref{fig:fig2}. The Workload Analyser Unit estimates future resource requirement on each server for successful execution of user requests by the respective virtual machines avoiding any performance degradation due to sudden upsurge or plunge in the forthcoming workload. This predicted resource usage information is passed to the resource management unit to provide it with the knowledge of the expected variance in the forthcoming workload and adjust the entire load with proficiency by handing all the under-utilization and over-utilization of resources proactively. 
 The resource management unit creates virtual networks of a group of active virtual machines by considering the trivial fact that all the virtual machines involved in execution of a common application are in communication, exchanging data with each other, are legally authorized to access data of one another. The resource management unit 208 also builds and maintains Authorized virtual machine (VM) Access Database (AVAD) by utilizing the link access information of aforesaid virtual networks of virtual machines. 
 A virtual machine (VM) Security Management Unit in accordance with the embodiment of the proposed model, which is capable of providing an effective security during execution of user application within the cloud infrastructure. It comprises of a monitor which observes live communication links among active virtual machine continuously in real-time. This monitor analyses virtual machine link access patterns wherein, the group of virtual machines involved in exchanging data among them generates live virtual networks of active virtual machines. The knowledge produced by these access links is utilized to create Online VM Access Log which is further utilized by the secure VM allocation unit which makes decisions of shifting, terminating, creating, re-allocating virtual machines in the favour of security of user data under processing.
 \par
  The virtual networks for live communication among virtual machines  are monitored  and recorded in a log named Current VM Access Links Log (CVAL). The information captured in this log about inter-communication among active virtual machines is analysed and compared with Authorised VM Access Database (AVAD) which is generated by resource management unit. All the mis-matched link are identified for detection and termination of unauthorized access links and attacker virtual machines prior to occurrence of successful security breaches. The present approach utilizes the knowledge produced by comparative analysis between AVAD and CVAL to furnish active virtual machines security decisions in real-time.
 
\section{Resource Management}

\begin{figure}[!htbp]
 	\centering
 	\includegraphics[width=1.0\linewidth]{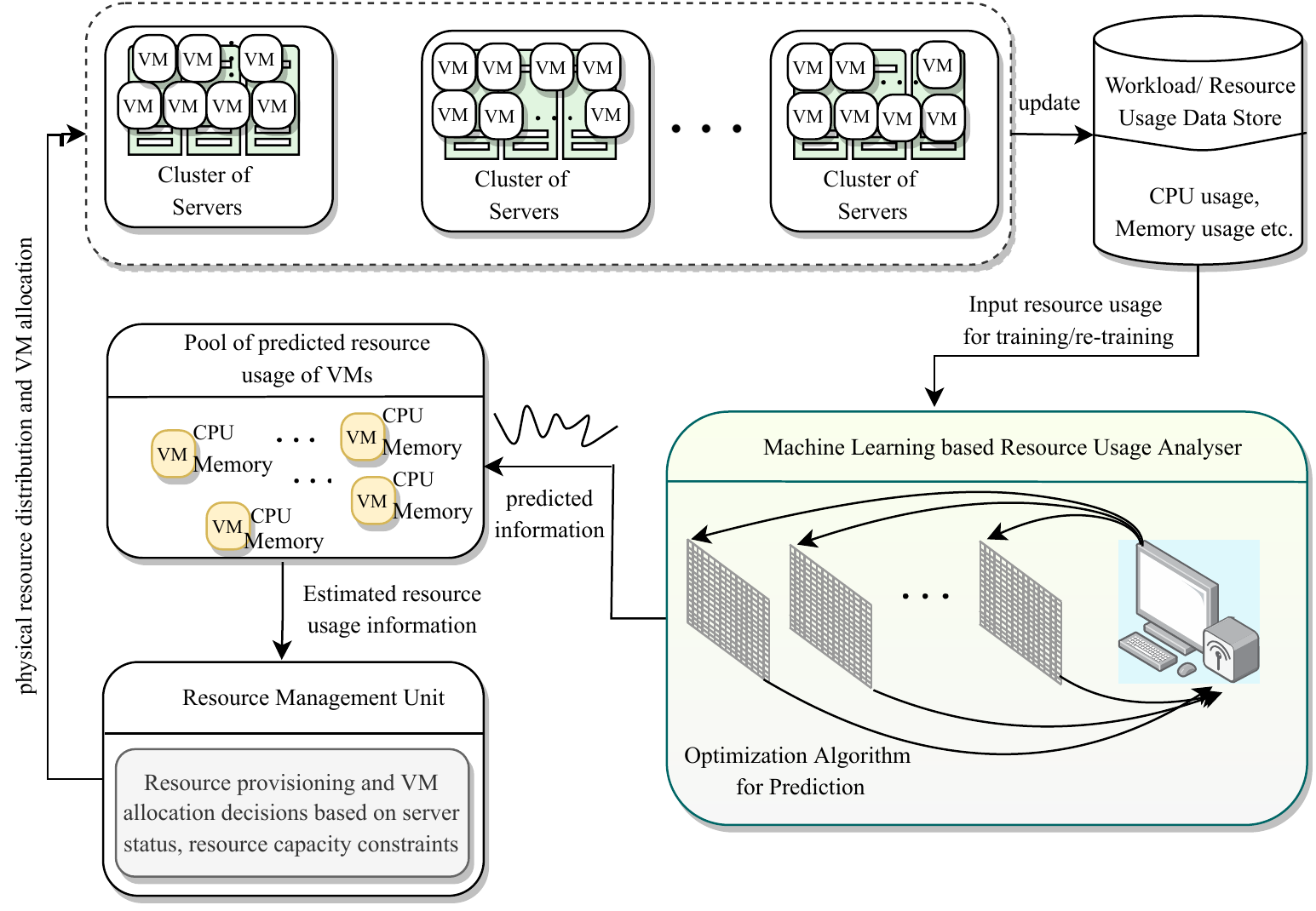}
 	\caption{Machine learning driven resource estimation unit}
 	\label{fig:fig2}
 \end{figure}

 A block diagram as illustrated in Fig. \ref{fig:fig2} is provided to elucidate an implementation of high performance and energy-proficient resource management in accordance with resource management unit presented in Fig. 2 of the proposed model. The historical and live resource utilization information from the cloud infrastructure  is stored in a resource usage data store. This is information is pre-processed by applying feature selection and normalization techniques which maybe performed using any appropriate machine learning method; such as Decision Tree, Boruta Forest, ChiSquare, or Random Forest etc. for feature selection and Linear scaling, Clipping, Max-Min algorithm, Z-score etc. for normalization of input data within a specific range. 
 The normalized resource usage data is provided to Machine Learning based Resource Usage Analyser  for training and re-training periodically in real-time. This analyser maybe developed using any of the prediction methods such as Neural Network perceptron trained with help of an optimization algorithm, Deep Neural Network, Long Short Term Memory based Recurrent Neural Network (LSTM-RNN) etc. The periodic training of the prediction module rejuvenates the prediction model for improved and precise prediction of the respective resource utilization within cloud infrastructure. This prediction information is generated in the form of patterns of resource usage within a virtual machine and thus produces a pool of predicted resource usage of VMs. This estimated resource usage information is passed to resource management unit which corresponds to resource management unit for directing resource provisioning decisions. 
 In case, the predicted capacity of resource usage is more than the physically available resource capacity of a server in the cluster, an overloading is expected. The highest resource capacity virtual machines are migrated from the respective server for alleviation of such an overloading problem that may degrade the performance of entire cloud infrastructure. The highest resource capacity virtual machines are selected for migration in order to avoid the necessity of frequent migration. Also, in case of much lesser predicted capacity of resource usage than the actual capacity of server, an underloading problem is estimated wherein, such an underloaded server maybe turned off after shifting the virtual machines hosted on them to some other active server subject to resource capacity constraints.

 \par
 The resource usage information from workload data store  maybe use for periodic training/re-training of machine learning based resource usage analyser for estimation of future workload and resource utilization. The information of inter-access link among active virtual machines in CVAL are compared with that of AVAD. If the outcome of the comparison block  is ‘NO’, any security breach is detected proactively which requires enough time for surpassing the security defence and actual occurrence of the successful attack and terminate the unauthorised links among respective VMs before initiation of data transmission, otherwise, no post-action is needed. Accordingly, the knowledge of predicted resource usage and unauthorized inter-communication links are passed to resource management unit that may allocate/deallocate or migrate virtual machines based on server resource capacity and security status of virtual machines. 
 
 \section{Illustrations}
 The effectiveness of the present approach against a co-residency attack is provided in accordance with an engagement of the proposed model. Let us assume that User 1 application’s component tasks are executing on multiple virtual machines, among which one of the virtual machines is co-located with an attacker’s virtual machine who may launch a side-channel or co-residency attack on all the virtual machines of User 1 by reason of networking links among them. Such an attack requires sufficient time for breaching the security boundaries of target virtual machine to launch a successful attack. The proposed model continuously monitors, records and analyses the inter-communication links among active virtual machines which successfully points out the unauthorized link as soon as it is developed. As depicted in proposed approach, the attacker virtual machine is detected and terminated providing successful defence against the security breach prior to its occurrence. \par
 The working strategy of the present approach against a multiple virtual machines hijack attack in accordance with an embodiment of the proposed model can be illustrated as follows. Suppose an attacker launched multiple virtual machines for maximum data breaches simultaneously by exploiting co-residency effect. Since, the proposed model continuously monitors, records and analyses the inter-communication links among active virtual machines which successfully points out the multiple unauthorized links as soon as they are developed. The VM Security Analysis unit detects and terminates all attacker virtual machines involved in data breaches via unauthorized access links before occurrence of successful attack. \par 
Furthermore, the virtual machine security mechanism of the present approach against a grouped virtual machines attack targeting a single virtual machine in accordance with the proposed model can be realised with the subsequent example. Consider an attacker launched multiple virtual machines to reach the target virtual machine via unauthorized link using network cascading effect. Since, the proposed model continuously monitors, records and analyses the inter-communication links among active virtual machines which successfully points out the multiple unauthorized links as soon as they are developed. The VM Security Analysis unit detects and terminates all attacker virtual machines involved in such an attack via unauthorized access links prior to its success.
\section{Conclusions} 
The proposed conceptual model for security and performance efficiency of cloud computing environment is a novel approach accomplishing two conflicting objectives concurrently. This paper presented an architectural and operational design of the proposed model with essential illustrations to validate the effectiveness of the proposed approach. In future, the authors envisions to implement this model and compare it with existing state-of-the-art approaches.  
 
\bibliographystyle{IEEEtran}
\bibliography{Main}

\end{document}